# Superconductivity in Metal Rich Li-Pd-B Ternary Boride

K. Togano[1], P. Badica[1], Y. Nakamori[1], S. Orimo[1], H. Takeya[2] and K. Hirata[2]

[1]Institute for Materials Research, Tohoku University, 2-1-1, Katahira, Aoba-ku, Sendai 980-8577, Japan

[2]National Institute for Materials Science, 1-2-1, Sengen, Tsukuba 305-0047, Japan

8K superconductivity was observed in the metal rich Li-Pd-B ternary system. Structural, microstructural, electrical and magnetic investigations for various compositions proved that $Li_2Pd_3B$ compound, which has a cubic structure composed of distorted $Pd_6B$ octahedrons, is responsible for the superconductivity. This is the first observation of superconductivity in metal rich ternary borides containing alkaline metal and Pd as a late transition metal. The compound prepared by arc melting has high density, is stable in the air and has an upper critical field, $H_{c2}(0)$, of 6T.

PACS numbers: 74.70.Dd, 74.25.Ha, 74.25.Fy, 74.25.Op

The search for superconductivity in boride compounds initiated in 1949 with the discovery of TaB ($T_c$=4K) [1]. However, most works were done in the 1970's and 1980's resulting in the discovery of many binary and ternary superconducting borides, almost all of which involve either transition metal elements, rare-earth elements and platinum group elements of Ru, Rh, Os, Ir and Pt as metal constituents [2]. Despite of these efforts, the transition temperature $T_c$ of binary and ternary borides remained below 12K and could not exceed 23K, which was the highest $T_c$ of intermetallic compounds recorded by A15 $Nb_3Ge$ in 1973 [3]. It is interesting that no Pd binary or ternary boride was reported in stable condition, although Pd belongs to platinum group and gives the highest $T_c$ in the case of quaternary borocarbide system (RE)(TM)$_2B_2C$, where RE=rare earth metal and TM=Ni,Pd or Pt [4].

Recent discovery of 39K superconductivity in $MgB_2$ [5] has led to resurgence of interest in boride compounds as possible high temperature superconductor. It is surprising that such high $T_c$ was attained for simple binary boride with light alkaline earth element, Mg, as a metallic constituent. Since then, there have been several experimental and theoretical studies to search for high $T_c$ borides extending the metallic constituent to alkaline and alkaline earth elements. In these, of particular interest was the prediction of high temperature superconductivity in $Li_{1-x}BC$ system [6], however, to date, no superconductivity was reported for this system [7].

In this paper, we report the discovery of superconductivity around 8K in a metal rich boride of $Li_2Pd_3B$ compound with cubic structure. This is the first report for superconductivity in ternary borides containing alkaline metal and Pd as a platinum group element and not containing any rare-earth and transition metal elements. The result is expected to provide a possible road for searching a new family of boride superconductors with high transition temperature.

The samples were prepared by arc-melting process in order to attain high-density material that is adequate to measure electrical properties. In order to minimize the loss of Li by evaporation, we employed a two-step arc melting process. Initially, the Pd-B binary alloy buttons were prepared by conventional arc melting method from the mixture of appropriate amounts of Pd (99.9%) and B (99.5 %). We prepared four alloys with different Pd:B ratios of 3:2, 5:2, 3:1 and 5:1. By this alloying, the melting point of the materials can be lowered below the boiling point of Li at 1atm. Weight loss during the first arc melting step, was negligible. The alloying of Li was done in the second arc melting processing step. A small block of Pd-B alloy (~200mg) obtained by crushing the button was placed on a small piece of Li plate (10-50mg) freshly cut from the Li ingot (>99%). The melting was done in ~1atm argon atmosphere and the arc current was controlled to the necessary minimum; once the Pd-B alloy melted, the reaction with Li occurred and developed very fast probably due to the self-heating generated in the exothermic-type reaction. The loss of Li was inevitable making difficult the control of Li concentration in the final products. Therefore, the Li concentration in the obtained Li-Pd-B alloy was estimated from the weight gain of the Pd-B alloy that is considered to have a constant weight during arc-synthesis.

Temperature dependence of magnetization

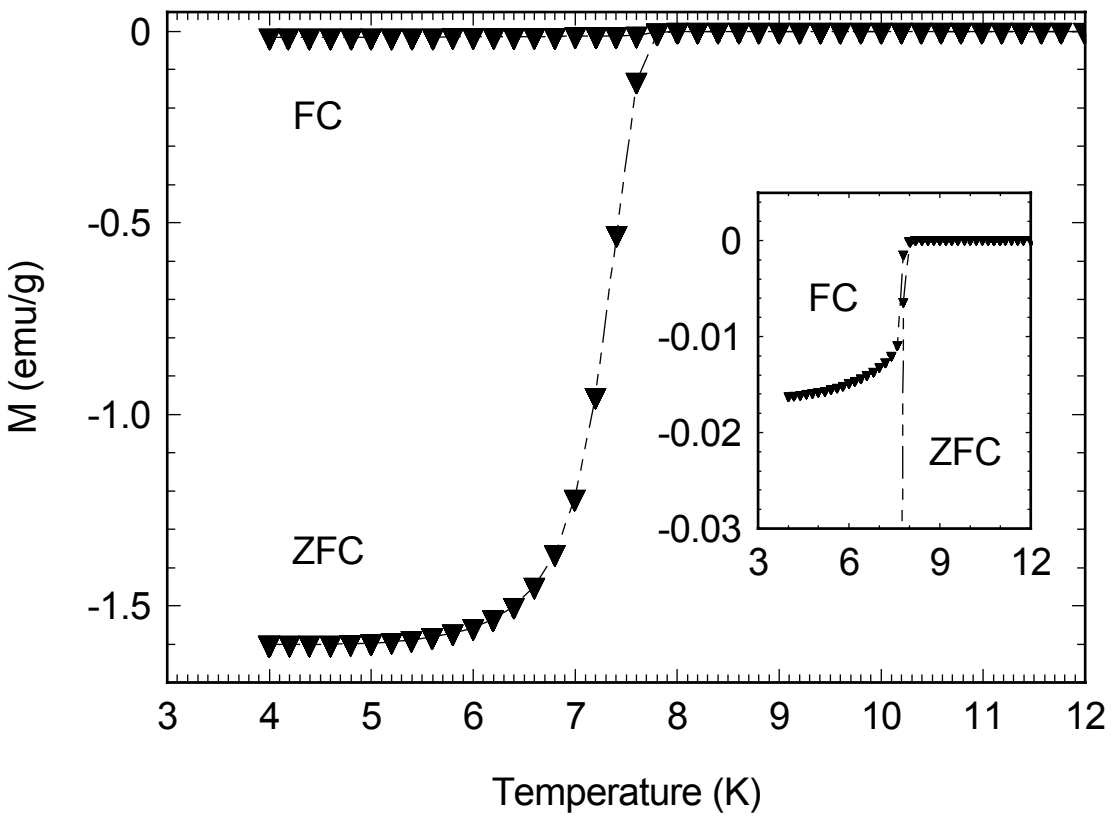

Figure 1 Magnetization vs. temperature curves for the $Li_2Pd_3B$ sample measured in zero-field-cooling (ZFC) and field-cooling (FC) arrangements in the magnetic field of 100Oe. Inset is showing the FC curve in detail.

was measured for the samples at 100Oe by a superconducting quantum interference device (SQUID) magnetometer. A sharp drop of magnetization at around 7-8K, which is the characteristic signature of superconductivity, was observed for a variety of compositions examined in this work. The largest diamagnetic signal (Fig. 1) was obtained for the sample with an estimated composition of approximately $Li_2Pd_3B$. The onset of the transition is 8K as shown in the inset of Fig. 1. The zero-field-cooling (ZFC) curve shows an almost full shielding effect, while the field-cooling (FC) curve shows a low Meissner effect of the order of 1% due to flux trapping. The sample has uniform solidification microstructure (Fig. 2), composed of grains with a few hundreds μm size with cellular dendrites inside the grain. Many cracks were observed along the grain boundaries and sub-grain boundaries. The powder X-ray diffraction pattern for this sample is shown in Fig. 3. All of the peaks can be ascribed to $Li_2Pd_3B$ compound, which was recently reported by Eibenstein and Jung [8]. The crystal structure is cubic and composed of distorted $Pd_6B$ octahedrons. No apparent peak of impurity phase was observed. The samples with estimated compositions different from $Li_2Pd_3B$ showed a broader transition with smaller diamagnetic signal, and at the same time extra peaks belonging to unidentified phases occurred in the XRD patterns. From these results, we conclude that the cubic $Li_2Pd_3B$ compound is responsible for the observed 8K- superconductivity. The sample is stable in the air and showed no significant difference of diamagnetic signal after 1 week.

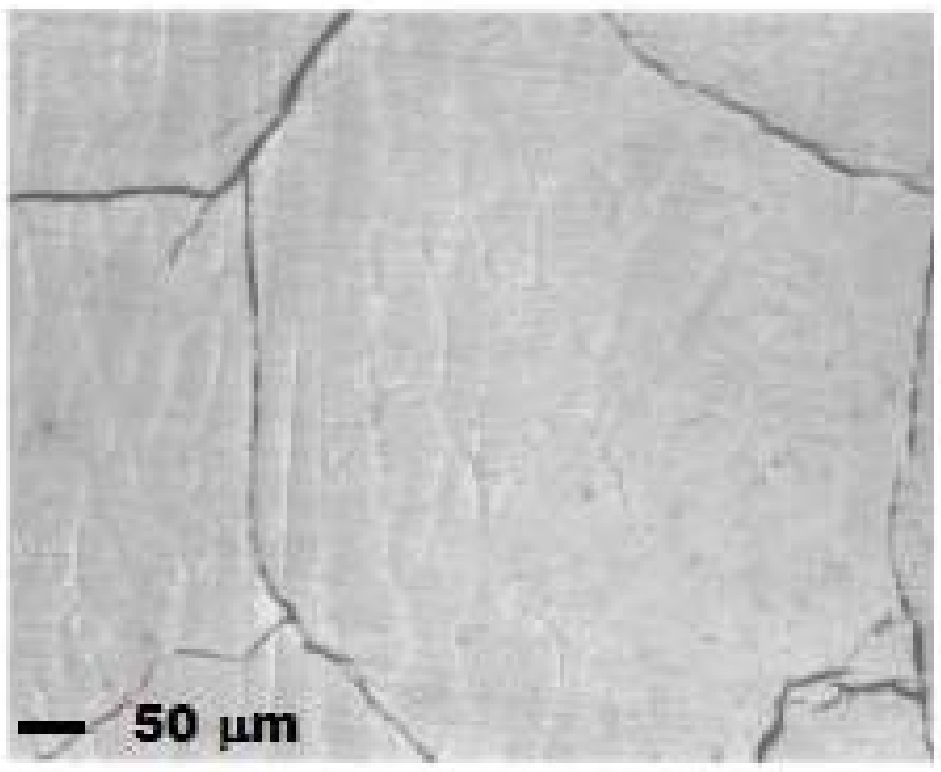

Figure 2 Optical microstructure on the cross section of the arc melted $Li_2Pd_3B$ button.

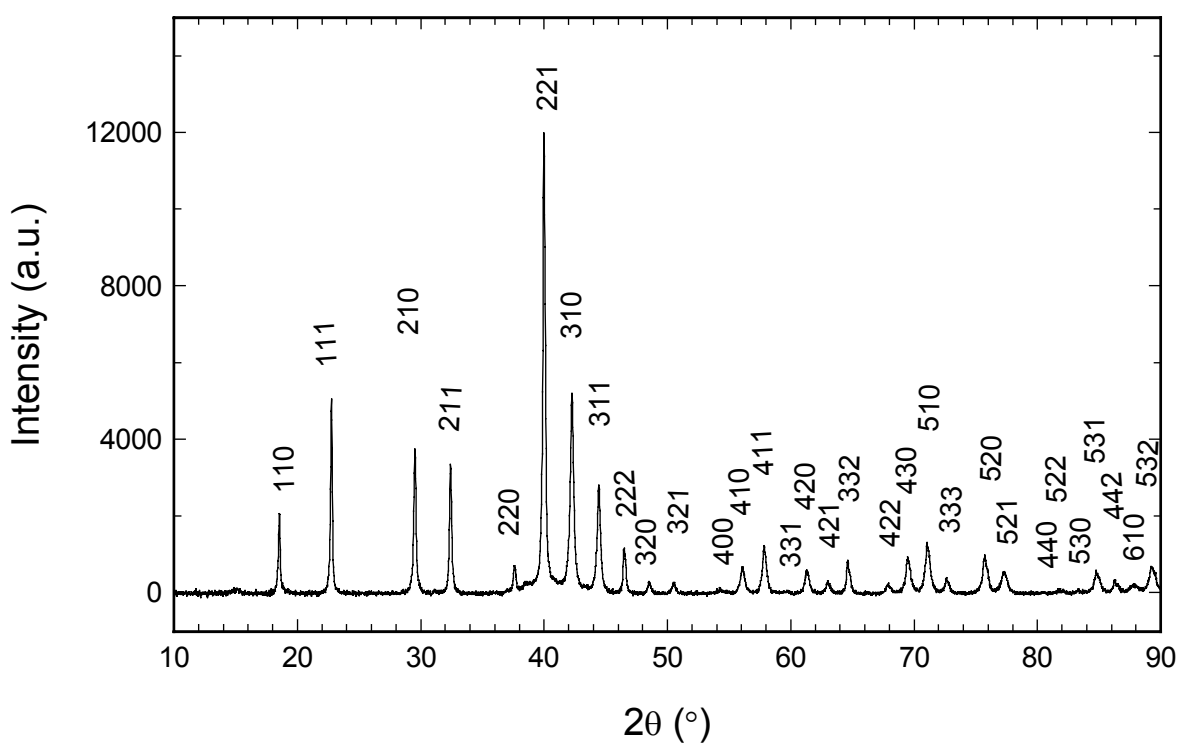

Figure 3 XRD powder pattern of the $Li_2Pd_3B$ sample synthesized by arc melting.

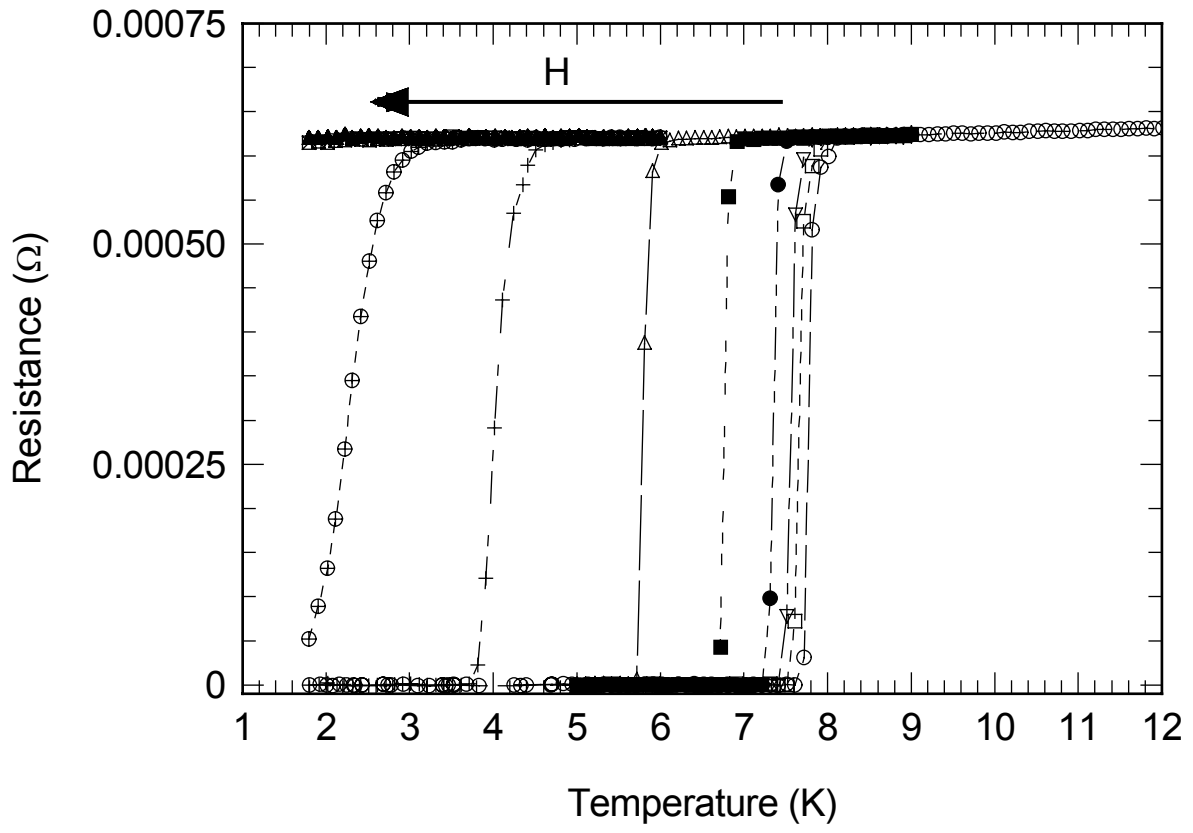

Figure 4 Temperature-dependent resistivity in applied magnetic fields of 0, 0.05, 0.1, 0.2, 0.5, 1, 2, 3, 4, 5T for the $Li_2Pd_3B$ bulk sample (a few mm size). Applied current was 1mA for a 4-probe configuration.

Figure 4 presents the temperature-dependent resistivity in applied magnetic fields up to 5T for $Li_2Pd_3B$ sample. The sample was a bulk (a few mm size) obtained by crushing the final arc melted product (button). Measuring applied current was 1mA. Under zero field, the curve drops sharply with the onset temperature of 8.2 K and the transition width of 0.6K. The onset temperature is slightly higher than that obtained from the magnetization measurement. In applied magnetic field, the curve shows the parallel shift to lower temperature, which is characteristic for the metallic superconductors. In Figure 5 is given the onset temperature as a function of magnetic field. The curve shows a positive curvature near to $T_c$ , similar to the polycrystalline borocabides [9] and $MgB_2$ [10]. Except this region, the curve is linear, whose gradient $dH_{c2}/dT$ is -0.84T/K. Linear

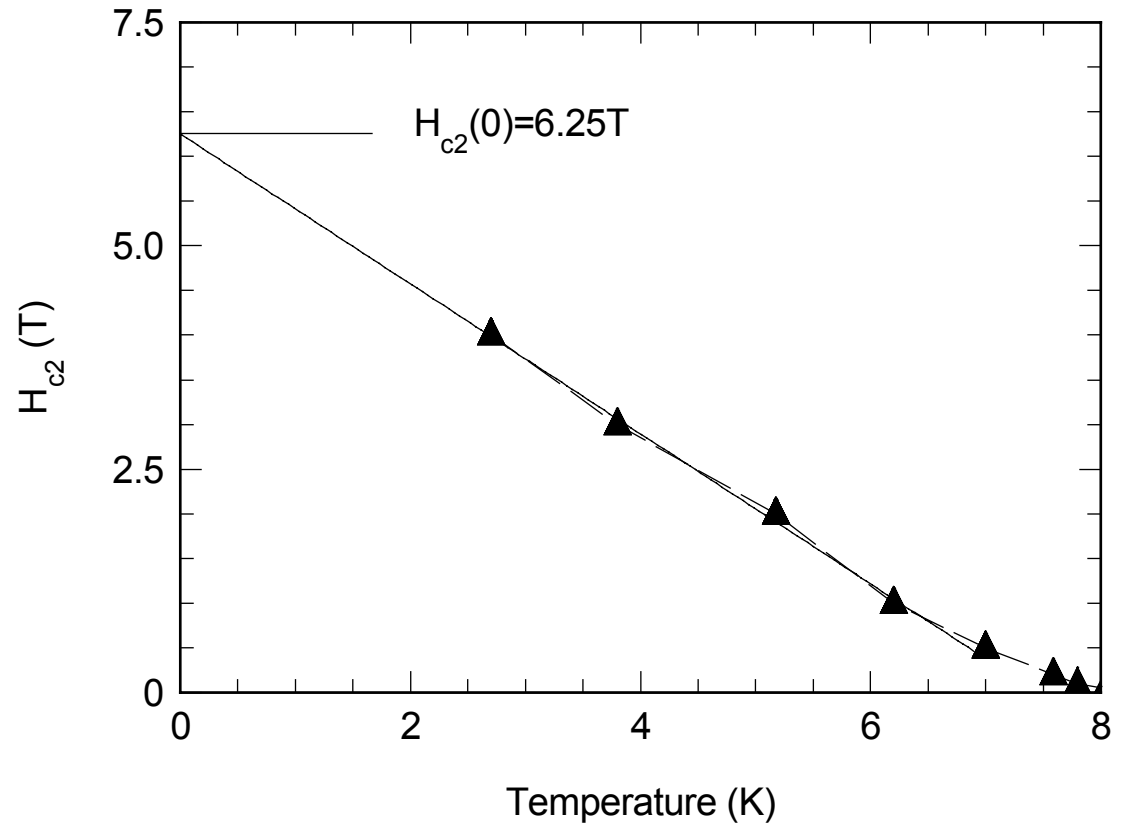

Figure 5 $H_{c2}(T)$ plot defined by the onset temperature of resistive transition measured in magnetic fields for the $Li_2Pd_3B$ sample.

extrapolation of the curve gives the upper critical field $H_{c2}(0)$ of 6.25T.

In summary, we found that the cubic $Li_2Pd_3B$ compound is a superconductor with critical temperature of about 8K and upper critical field $H_{c2}(0)$ of 6T. The compound prepared by arc-melting has a uniform structure and high density, and is stable in the air. This is the first observation of superconductivity in ternary metal rich borides composed of alkaline metal, platinum group element and boron. The result is expected to provide a new direction for searching new types of boride superconductors with high superconducting transition temperature.

The authors would like to thank Mr. T. Kondo and Mr. T. Kudo for technical assistance during experiments. The study was carried out as a part of "Ground-based Research Announcement for Space Utilization" promoted by Japan Space Forum.